\documentstyle[12pt]{article}
\def\ftoday{{\sl  \number\day \space\ifcase\month 
\or Janvier\or F\'evrier\or Mars\or avril\or Mai
\or Juin\or Juillet\or Ao\^ut\or Septembre\or Octobre
\or Novembre \or D\'ecembre\fi
\space  \number\year}}    

\makeatletter
\@addtoreset{equation}{section}
\makeatother

\newcommand{\es}{\\[3mm]}

\renewcommand{\a}{\alpha}
\renewcommand{\b}{\beta}
           
\renewcommand{\d}{\delta}         
\newcommand{\e}{\varepsilon}

\newcommand{\m}{\mu}
\newcommand{\n}{\nu}

\newcommand{\th}{\theta}         
           
\newcommand{\vf}{{\varphi}}
\newcommand{\x}{\xi}

\newcommand{\OO}{{\cal O}}

\newcommand{\Sla}{\raise.15ex\hbox{$/$}\kern -.70em D}

\newcommand{\lp}{\left(}\newcommand{\rp}{\right)}
\newcommand{\lc}{\left[}\newcommand{\rc}{\right]}
\newcommand{\lac}{\left\{}\newcommand{\rac}{\right\}}

\newcommand{\complex}{{\kern .1em {\raise .47ex
\hbox {$\scriptscriptstyle |$}}
    \kern -.4em {\rm C}}}
\newcommand{\real}{{{\rm I} \kern -.19em {\rm R}}}
\newcommand{\rational}{{\kern .1em {\raise .47ex
\hbox{$\scripscriptstyle |$}}
    \kern -.35em {\rm Q}}}
\renewcommand{\natural}{{\vrule height 1.6ex width
.05em depth 0ex \kern -.35em {\rm N}}}

\newcommand{\tr}{{\rm {Tr} \,}}

\newcommand{\pa}{\partial}

\newcommand{\fud}[2]{{\frac{\delta #1}{\delta #2}}}

\newcommand{\ie}{{{\em i.e.},\ }}

\newcommand{\sla}{\raise.15ex\hbox{$/$}\kern -.57em}
\newcommand{\twiddle}{\lower.9ex\rlap{$\kern -.1em\scriptstyle\sim$}}

\newcommand{\vev}[1]{\left\langle {#1}\right\rangle}
\newcommand{\equ}[1]{(\ref{#1})}
\newcommand{\eq}{\begin{equation}}
\newcommand{\eqn}[1]{\label{#1}\end{equation}}
\newcommand{\eea}{\end{eqnarray}}
\newcommand{\eqa}{\begin{eqnarray}}
\newcommand{\eqan}[1]{\label{#1}\end{eqnarray}}
\newcommand{\ba}{\begin{array}}
\newcommand{\ea}{\end{array}}
\newcommand{\eqac}{\begin{equation}\begin{array}{rcl}}
\newcommand{\eqacn}[1]{\end{array}\label{#1}\end{equation}}

\newcommand{\nes}{\nonumber \\[3mm]}
\renewcommand{\title}[1]{\null\vspace{25mm}

\noindent{\Large{\bf #1}}\vspace{10mm}

}
\newcommand{\authors}[1]{\noindent{\large #1}\vspace{20mm}

}
\newcommand{\address}[1]{{\center{\noindent #1\vspace{10mm}}

}}
\renewcommand{\abstract}[1]{\vspace{17mm}

\noindent{\small{\em Abstract.} #1}\vspace{2mm}

}   
\begin{document} 
\begin{titlepage}
\begin{center}
\hspace*{\fill}{{\normalsize \begin{tabular}{l}
                              {\sf  UGVA---DPT 1996/09--952}\\
                              {\sf TUW 96-19}\\
                              {\sf hep-th/9609240}
                         \end{tabular}   }}

\title{Two--Dimensional BF Model \es
        Quantized in the Axial Gauge}

\authors{S.~Emery$^{1\ \vdash}$, M.~Kr\"uger$^\dashv$, J.~Rant$^\dashv$, 
         M.~Schweda$^\dashv$ and T.~Sommer$^{2\ \dashv}$}
       
\address{$^\vdash$ D\'epartement de Physique Th\'eorique, 
                          Universit\'e de Gen\`eve\\
       24 quai E. Ansermet, CH -- 1211 Gen\`eve 4 (Switzerland)\es
        $^\dashv$ Institut f\"ur Theoretische Physik, 
                          Technische Universit\"at Wien\\
      Wiedner Hauptstra\ss e 8-10, A-1040 Wien (Austria)}
        
\end{center}
\footnotetext[1]{Supported in part by the Swiss National Science 
                       Foundation.}
\footnotetext[2]{Supported in part by the FWF, project 
        no. P-10268-PHY}
\abstract{
The two--dimensional topological BF model is quantized in the axial gauge.
We show that this theory is trivially ultraviolet finite and that the 
usual infrared problem of the propagator of the scalar field in two 
dimensions is replaced by an easily solvable long distances problem inherent
to the axial gauge. It will also be shown that contrarily to the 
3--dimensional case, the action principle 
cannot be completely replaced by the various Ward identities expressing 
the symmetries of the model; some of the equation of motion are needed. 
}

\thispagestyle{empty}
\end{titlepage}

\section{Introduction}

The main property of Topological Field Theories (TFT) \cite{topol} is the 
fact that 
their observables are of topological nature, i.e. they only depend on the 
global properties of the manifold on which the theory is defined. 
Therefore, these kind of theories exhibit remarkable ultraviolet 
finiteness properties due to the lack of any physical, {\it i.e.} metric 
dependent observables. In particular, the two--dimensional BF model 
treated here provides an example of a fully finite quantum field theory.
 
The history of TFT's is closely related to 
the relationship between the problems that arise in the research of 
physical systems and the ensuing mathematical methods that are needed for
their solution. 
A well known example is the relation between the work of Donaldson
concerning the study of the topology of four
dimensional manifolds \cite{donaldson1,donaldson2} and its description 
as a Topological Yang--Mills theory due to Witten \cite{yangmills}.
A further example is the study of the knot and 
link invariants in the case of three--dimensional Chern--Simons theory 
\cite{witten2}. 

The Topological Yang--Mills theory and the Chern--Simons theory are 
examples of two distinct classes of TFT's, the  
former belonging to the Witten type -- 
{\it i.e.} the whole action is a BRST variation -- and 
the latter being of the Schwarz type --
the action splits into an invariant part and a BRST variation term. 

There exists a further type of Schwarz class TFT's: the Topological 
BF models \cite{bfmod}. They constitute the natural extension of the 
Chern--Simons model in an arbitrary number of spacetime dimensions.
These models describe the coupling of an antisymmetric 
tensor field to the Yang--Mills field strength \cite{antisym}, 
\cite{blau}.

The aim of the present work is the quantization of the two--dimensional 
BF model in the axial gauge. 
It is motivated by similar works done for the Chern--Simons model 
\cite{chern3} and the BF model \cite{bf3} both in three spacetime dimensions. 
The axial gauge is particularly interesting for these two models since in 
this gauge these two theories are obviously ultraviolet 
finite due to the complete absence of radiative 
corrections. Since for the two--dimensional case we are interested in,
the choice of the axial gauge allows us to overcome the usual infrared
problem which occurs in the propagator of the dimensionless 
fields \cite{blasi2}, it is interresting to see wether the ultraviolet
finitness properties are also present.

On the other hand, we have already shown that for the three--dimensional 
Chern-Simons and BF models \cite{chern3}, \cite{bf3}, the symmetries 
completely define  the theory, {\it i.e.} the quantum action principle
is no more needed. The latter property relies on the existence of a 
topological linear vector supersymmetry \cite{delducgs} besides the BRST 
invariance \cite{brst1,brst2}. The generators of the latter together with 
the one of the BRST--symmetry form a superalgebra 
of the Wess--Zumino type \cite{wesszum1,wesszum2}, which closes 
on the translations \cite{sor}.  Then the associated Ward identities 
can be solved for the Green's functions, exactly and uniquely.

As this linear vector supersymmetry is also present in two--dimensions,
it is natural to ask ourself wether all the Green's functions of the 
two--dimensional model are also uniquely determined by symmetry
considerations. We will answer this question by the negative. The 
consistency relations let some class of Green's functions undetermined 
and therefore one has to use some of the equations of motion.

The present work is organized as follows.  The model is introduced in 
section \ref{model} and its symmetries are discussed. In section 
\ref{symdisc} we will check the consistency conditions between the various 
symmetries and thus recover almost all the equations of motion 
for the theory. Section \ref{prop_finit} is devoted to the derivation of 
the Green's functions keeping in mind that some are consequences of 
symmetries only whereas some others are solution of the field equations. 
 At the end we draw some conclusions.

\section{The two--dimensional BF model in the axial gauge} \label{model}

The classical BF model in two spacetime dimensions
is defined by 
\begin{equation} \label{s_inv}
    S_{\mbox{\small inv}} = \frac{1}{2} \,\int\limits_{\cal{M}} d^2 x \, 
    \varepsilon^{\mu \nu} F^a_{\mu \nu} \phi^a \quad .
\end{equation}
One has to stress that the action (\ref{s_inv}) does not depend on a 
metric $g_{\mu \nu}$ which one may introduce on the arbitrary 
two--dimensional manifold ${\cal M}$.
In this paper, ${\cal M}$ is chosen to be the flat Euclidean 
spacetime\footnote{${\cal M}$ has the metric 
$\eta_{\mu \nu} = \mbox{diag}\, (+\, 1 , +\, 1)$.}.
$\varepsilon^{\mu \nu}$ is the totally antisymmetric 
Levi--Civita--tensor\footnote{
The tensor $\varepsilon^{\mu \nu}$ is normalized to $\varepsilon^{12} =
\varepsilon_{12} = 1$ and one has 
$\varepsilon^{\mu \nu} \varepsilon_{\sigma\tau} = 
\delta^\mu_\sigma \delta^\nu_\tau - \delta^\mu_\tau \delta^\nu_\sigma$.}.  
The field strength $F^a_{\mu \nu}$ is related to the Yang--Mills gauge 
field $A^a_\mu$ by the structure equation
\begin{equation}
        F^a_{\mu \nu} = \partial_\mu A^a_\nu - \partial_\nu A^a_\mu
        + f^{abc} A^b_\mu A^c_\nu \quad ,
\end{equation}
and the $B$ field, traditionally denoted by $\phi$ for the two
dimensional case, is a scalar field. 

The fields ($A^a_\mu$, $F^a_{\mu \nu}$, $\phi^a$) belong to the adjoint 
representation of a simple compact gauge group ${\cal G}$.
The corresponding generators $T^a$ obey
\begin{equation}
   \big[ T^a , T^b \big] = f^{abc} T^c \quad , 
     \quad Tr(T^a T^b) = \delta^{ab}
        \quad ,
\end{equation}
$f^{abc}$ are the totally antisymmetric structure constants 
of ${\cal G}$. The field--strength satisfies the Bianchi--identity
\begin{equation} \label{bianchi1}
        (D_\rho F_{\mu \nu})^a + (D_\mu F_{\nu \rho})^a + 
        (D_\nu F_{\rho \mu})^a = 0 \quad ,
\end{equation}
where the covariant derivative $D_\mu$ in the adjoint 
representation is
\begin{equation} \label{covariant_der}
        (D_\mu \,\cdot \, )^a = (\partial_\mu \, \cdot \, )^a 
              + f^{abc} A^b_\mu (\, \cdot \, )^c \quad .
\end{equation}

The equations of motion are
\begin{eqnarray}
  \label{eomac} \frac{\delta \, S_{\mbox{\small inv}}}{\delta \,\phi^a}  
                      & = & \frac{1}{2}\, \varepsilon^{\mu \nu} 
                             F^a_{\mu \nu} \; = \; 0 \quad , \\
  \label{eomphi} \frac{\delta \, S_{\mbox{\small inv}}}{\delta \, A^a_\mu} 
             & = & \varepsilon^{\mu \nu} (D_\nu \phi)^a \; = \; 0 \quad .
\end{eqnarray}
Eq. (\ref{eomac}) implies the vanishing curvature condition,
\begin{equation}
        F^a_{\mu \nu} = 0 \quad ,
\end{equation} 
and from eq. (\ref{eomphi}) follows that the scalar field is confined on 
the hypersphere 
\begin{equation}
        \phi^a \phi^a = \mbox{const.} \quad .
\end{equation}

The action (\ref{s_inv}) is invariant under the infinitesimal gauge 
transformations 
\eq\ba{rcl}
     \label{deltaa} \delta A^a_\mu & = & -\, (\partial_\mu \theta^a + 
        f^{abc} A^b_\mu \theta^c) \; = \; -\, (D_\mu \theta)^a \quad , \es
     \label{deltaphi} \delta \phi^a & = & f^{abc} \theta^b \phi^c \quad ,
\ea\eqn{gautrans}
where $\theta^a$ is the infinitesimal local gauge parameter.

We choose to work in the axial gauge
\begin{equation} \label{axgauge}
n^\mu A^a_\mu =0\, .
\end{equation}
 and without loss of generality, we fix the axial gauge vector to be
\eq
        n^\mu = ( n^1 , n^2 ) = ( 1 , 0 ) \quad .
\eqn{axdef}
Following the usual BRST procedure, one introduces
a ghost, an antighost and a Lagrange multiplier field 
($c^a$, $\bar{c}^a$, $b^a$) in order to construct the gauge fixed 
action $S$ in a manifestly BRST--invariant manner,
\begin{equation} \label{total_s}
        S = S_{\mbox{\small inv}} + S_{\mbox{\small gf}} \quad ,
\end{equation}
\begin{eqnarray}
 S_{\mbox{\small gf}} & = & \int d^2 x \,  s (\bar{c}^a n^\mu 
  A^a_\mu ) \nonumber \\
  & = & \int d^2 x \,( b^a  n^\mu A^a_\m + \bar{c}^a n^\mu\partial_\m 
  c^a + \bar{c}^a  f^{abc} n^\mu A^b_\m c^c) \quad .
\end{eqnarray}
The nilpotent BRST--transformations are given by
\begin{equation} \label{brst_transf} 
\begin{array}{rcl}
        s A^a_\mu & = & -\,(D_\mu c)^a \quad , \\[2mm]
        s \phi^a & = & f^{abc} c^b \phi^c \quad , \\[2mm]
        s c^a & = & \frac{1}{2} \, f^{abc} c^b c^c \quad , \\[2mm]
        s \bar{c}^a & = & b^a \quad ,\\[2mm]
        s b^a & = & 0 \quad ,  \\[2mm]
        s^2 & = & 0 \quad .
\end{array}
\end{equation}  
Note that the gauge fixing term $S_{\mbox{\small gf}}$ is not metric 
independent and therefore not topological.

The canonical dimensions and ghost charges of the fields are collected in 
table \ref{table1}, where the canonical dimension of the gauge direction 
$n^\mu$ is zero.
\begin{table}[h] 
\begin{center}
\begin{tabular}{|l||c|c|c|c|c|c|} \hline
        & $A^a_\mu$ & $\phi^a$ & $c^a$ & 
        $\bar{c}^a$ & $b^a$ \\ \hline \hline
        canonical dimension & 1 & 0 & 0 & 1 & 1 \\ \hline
        ghost charge ($Q_{\Phi\Pi}$) & 0 & 0 & 1 & -1 & 0 \\ 
        \hline
\end{tabular}
\end{center}
\caption{\label{table1} Canonical dimension and ghost charge of 
the fields}
\end{table}

Before investigating further let us make some comments concerning the
axial gauge. It is well known that this choice does not
fix the gauge completely. Indeed (\ref{total_s}) is still invariant under 
gauge transformations of the same type as \equ{gautrans} but where
the gauge parameter $\th^a$ depends only on  
$x^1$. This residual gauge invariance will play an important 
role in the sequel.

The action (\ref{total_s}) possesses, besides the BRST--symmetry 
(\ref{brst_transf}) and the scale invariance,  an 
additional linear vector supersymmetry. In order to derive the vector 
supersymmetry transformations, one considers 
the energy--momentum tensor $T_{\mu \nu}$ of the theory.
Since the invariant part of the action (\ref{s_inv}) is 
metric--independent, the improved energy--momentum tensor is an exact 
BRST--variation,
\begin{equation} \label{tmunu}
        T_{\mu \nu} = s \Lambda_{\mu \nu} \quad ,
\end{equation}
where $\Lambda_{\mu \nu}$ is given by
\begin{equation} \label{lmunu}
  \Lambda_{\mu \nu} = \eta_{\mu \nu} \,( \bar{c}^a \, n^\lambda 
  A^a_\lambda) - \bar{c}^a \,(n_\mu A^a_\nu + n_\nu A^a_\mu) \quad .
\end{equation}
Eq.~(\ref{tmunu}) explicitly shows the unphysical character of the
topological BF model. 
Using the functional form for the equations of motion 
\begin{eqnarray}
  \label{eoma} \frac{\delta S}{\delta A^a_\nu} & = 
               & \varepsilon^{\nu \mu} D_\mu
              \phi^a + n^\nu (b^a - f^{abc} \bar{c}^b c^c) \quad , \\
  \label{eomf} \frac{\delta S}{\delta \phi^a} & = & \frac{1}{2} 
             \,\varepsilon^{\mu \nu} F^a_{\mu \nu} \quad , \\
 \label{eomc}  \frac{\delta S}{\delta c^a} & =
         & n^\mu (D_\mu \bar{c})^a \quad ,  \\
 \label{eomcb}  \frac{\delta S}{\delta \bar{c}^a} & = 
        & n^\mu (D_\mu c)^a \quad ,
\end{eqnarray} 
and the gauge condition 
\begin{equation} \label{gaucond}
        \frac{\delta S}{\delta b^a} = n^\mu A^a_\mu \quad ,
\end{equation}
one gets for the divergence of (\ref{lmunu}) 
\begin{equation}
        \partial^\nu \Lambda_{\nu \mu} = \varepsilon_{\mu \nu} n^\nu
                \bar{c}^a \, \frac{\delta S}{\delta \phi^a} - A^a_\mu \,  
                \frac{\delta S}{\delta c^a} + \partial_\mu \bar{c}^a \, 
                \frac{\delta S}{\delta b^a} + \mbox{total deriv.} \quad . 
\end{equation}
An integration over the two--dimensional spacetime yields the 
Ward--identity of the linear vector supersymmetry,
\begin{equation} \label{wi_s}
        {\cal W}_\mu S = 0 \quad ,
\end{equation}
where
\begin{equation} \label{wi_op}
        {\cal W}_\mu = \int d^2x \,\Big( \varepsilon_{\mu \nu} n^\nu
                \bar{c}^a \, \frac{\delta}{\delta \phi^a} - A^a_\mu \,  
                \frac{\delta}{\delta c^a} + \partial_\mu \bar{c}^a \, 
                \frac{\delta}{\delta b^a} \Big) \quad 
\end{equation}
and  the transformations read
\begin{eqnarray} \label{susy_on}
        \delta_\mu A^a_\nu & = & 0 \quad , \nonumber \\ 
        \delta_\mu \phi^a & = & \varepsilon_{\mu \nu} n^{\nu} 
                \bar{c}^a \quad , \nonumber \\
        \delta_\mu c^a & = & - \, A^a_\mu \quad , \nonumber \\
        \delta_\mu \bar{c}^a & = & 0 \quad , \nonumber \\
        \delta_\mu b^a & = & \partial_\mu \bar{c}^a \quad .
\end{eqnarray}

{\em Remark:} It can be easily verified that $S_{\mbox{\scriptsize inv}}$ 
and $S_{\mbox{\scriptsize gf}}$ are not separately invariant under 
(\ref{susy_on}); only the combination (\ref{total_s}) is invariant. 

The generator $\delta_\mu$ of the linear vector supersymmetry and the 
BRST--operator $s$ form a graded algebra of the Wess--Zumino type which 
closes on--shell on the translations,
\begin{eqnarray}
\big\{ s , \delta_\mu \big\} A^a_\nu 
            & = & \partial_\mu A^a_\nu - \varepsilon_{\mu \nu} \,
              \frac{\delta S}{\delta \phi^a} \quad , \nonumber \\
\big\{ s , \delta_\mu \big\} \phi^a 
            & = & \partial_\mu \phi^a + \varepsilon_{\mu \nu} \,
              \frac{\delta S}{\delta A^a_\nu} \quad , \nonumber \\
\big\{ s , \delta_\mu \big\} \psi^a & = & \partial_\mu \psi^a \quad , 
         \quad \forall \,\psi^a \in \{ c^a , \bar{c}^a , b^a \} \quad .
\end{eqnarray}
Moreover, the following algebraic relations hold,
\begin{equation}
        \big\{ \delta_\mu , \delta_\nu \big\} = 0 \quad , \quad
        \big[ \delta_\mu , \partial_\nu \big] = 0 \quad .
\end{equation}

\section{Consequences of the symmetries} \label{symdisc}

We will now discuss the implications of the
symmetries discussed above independently of the action
(\ref{total_s}). This means that we are going to consider only
the gauge fixing condition, the Ward identity for the vector 
supersymmetry and the Slavnov--Taylor identity and look for 
their consistency independently from the field equations 
(\ref{eoma})--(\ref{eomcb}). Since we are ultimately interested in 
deriving all the Green's functions of the theory, 
let us first  rewrite all these functional identities in 
term of the generating functional for the connected Green's functions. 
We will see later that due to the absence of loop graphs for this theory,  
it is sufficient to restrict ourself to
the tree approximation,  {\it i.e.} the Legendre transformation of the 
classical action: 
\begin{eqnarray}
\frac{\delta Z^{\mbox{\small c}}}{\delta j^{\m \,a}} \, = \, A^a_\mu  
    & \quad , \quad &
     \frac{\delta S}{\delta A^a_\mu} \, = \, -\,j^{\m \,a} \quad , \\
\frac{\delta Z^{\mbox{\small c}}}{\delta j^a_\phi} \, = \, \phi^a 
    & \quad , \quad &
     \frac{\delta S}{\delta \phi^a} \, = \, -\,j^a_\phi \quad , \\
\frac{\delta Z^{\mbox{\small c}}}{\delta j^a_c} \, = \, c^a 
    & \quad , \quad &
     \frac{\delta S}{\delta c^a} \, = \,\hspace{4mm} j^a_c \quad , \\
\frac{\delta Z^{\mbox{\small c}}}{\delta j^a_{\bar{c}}} \, = \, \bar{c}^a 
    & \quad , \quad &
     \frac{\delta S}{\delta \bar{c}^a} \, 
                 = \,\hspace{4mm} j^a_{\bar{c}} \quad , \\
\frac{\delta Z^{\mbox{\small c}}}{\delta j^a_b} \, = \, b^a 
    & \quad , \quad &
      \frac{\delta S}{\delta b^a} \, = \, -\,j^a_b \quad , 
\end{eqnarray}
\begin{eqnarray} 
\label{legendre}
\lefteqn{Z^{\mbox{\small c}}[j^{\m \,a}, j^a_\phi, j^a_c, j^a_{\bar{c}}, 
j^a_b] =  S[A^a_\mu, \phi^a, c^a, \bar{c}^a, b^a] +} \\
         &&  +\, \int d^2x \,\Big( j^{\m \,a} A^a_\mu + 
              j^a_\phi \phi^a + j^a_c c^a + j^a_{\bar{c}} \bar{c}^a + 
              j^a_b b^a \Big) \quad , \nonumber
\end{eqnarray}
where the canonical dimensions and ghost charges of the classical 
sources $j^{\m \,a}$, $j^a_\phi$, $j^a_c$, $j^a_{\bar{c}}$, $j^a_b$ are 
given in table \ref{table2}. 
\begin{table}[h]
\begin{center}
\begin{tabular}{|l||c|c|c|c|c|} \hline
     & $j^{\m \,a}$ & $j^a_\phi$ & $j^a_c$ & $j^a_{\bar{c}}$ & $j^a_b$ \\ 
        \hline \hline
        canonical dimension & 1 & 2 & 2 & 1 & 1 \\ \hline
        ghost charge ($Q_{\Phi \Pi}$) & 0 & 0 & -1 & 1 & 0 \\ \hline 
\end{tabular}
\end{center} 
\caption{\label{table2} Dimensions and ghost charges of the sources}
\end{table}

The gauge condition (\ref{gaucond}) now reads
\begin{equation}
\label{gauge_cond}
n^\m \frac{\delta Z^{\mbox{\small c}}}{\delta j^{\m \,a}}
=-\, j^a_b \quad ,
\end{equation}
and the linear vector supersymmetry Ward identity (\ref{wi_s}) 
\begin{equation}
{\cal W}_\mu Z^{\mbox{\small c}} = 
   \int d^2x \,\Big[ -\,\varepsilon_{\mu \nu } n^\nu j^a_\phi 
   \,\frac{\delta Z^{\mbox{\small c}}}{\delta j^a_{\bar{c}}} -
 j^a_c \,\frac{\delta Z^{\mbox{\small c}}}{\delta j^{\m \,a}} 
  - j^a_b \,\partial_\mu
  \frac{\delta Z^{\mbox{\small c}}}{\delta j^a_{\bar{c}}} \Big] 
     \; = \; 0 \quad .
\end{equation}
The BRST--invariance is formally expressed by the Slavnov--Taylor 
identity
\begin{eqnarray} \label{stid}
    {\cal S}(Z^{\mbox{\small c}}) & = & \int d^2x \, \Big( j^{\m \,a} 
    \big[ (D_\mu c)^a \big] \cdot Z^{\mbox{\small c}} - j^a_\phi 
    \big[ f^{abc} c^b \phi^c \big] \cdot Z^{\mbox{\small c}}+\nonumber \\
 &&\hspace{1.7cm} +\, j^a_c \big[ \frac{1}{2} f^{abc} c^b c^c \big]\cdot 
        Z^{\mbox{\small c}} 
  + j^a_{\bar{c}} \frac{\delta Z^{\mbox{\small c}}}{\delta j^a_b} \Big) 
        \; = \; 0 \quad .
\end{eqnarray}
We have used the notation  $[\OO ] \cdot Z^c$ for the generating functional 
of the connected Green's functions with the insertions of a local field 
polynomial operator  $\OO $. Usually, such insertions must be renormalized 
and their renormalization is controlled by coupling them to external 
sources. But in the case of the axial gauge, these insertions are trivial 
due to the fact that the ghost fields decouple from the gauge field as 
we will see later.

In order to analyze the consequences of these functional identities, let 
us begin by the projection of the supersymmetry Ward--identity  
along the axial vector $n^\mu$ 
\begin{equation} \label{n_wi_z}
        n^\mu {\cal W}_\mu Z^{\mbox{\small c}} = 
        -\int d^2x \, j^a_b 
                \,\Big(\underbrace{-\, j^a_c + n^\mu \partial_\mu 
         \frac{\delta Z^{\mbox{\small c}}}{\delta j^a_{\bar{c}}}}_{X^a}
                \Big) = 0 \quad ,
\end{equation}
where the gauge condition (\ref{gauge_cond}) has been used.
Locality, scale invariance and ghost charge conservation imply that
$X^a$ is a local polynomial in the classical sources $j$ and their
functional derivatives $\delta/\delta j$ of dimension $2$ and ghost
charge $-1$. The most general form for $X^a$ may depends on a further
term
\eq
X^a =-j^a_c+n^\mu\partial_\mu\frac{\delta Z^{\mbox{\small c}}}
{\delta j^a_{\bar{c}}}+z^{abc}j^b_b\frac{\delta Z^{\mbox{\small c}}}
{\delta j^c_{\bar{c}}} \quad ,
\eqn{Joe1}
provided $z^{abc}$ is antisymmetric in $a$ and $b$. The latter is thus
 proportional to the structure constants $f^{abc}$.
By substituting the general form (\ref{Joe1}) into (\ref{n_wi_z})
\begin{equation}
\int d^2x \, j^a_b\,X^a=0 \quad,
\end{equation}
one gets
\begin{equation}
\label{antighost_z}
        n^\mu \partial_\mu \Big( \frac{\delta Z^{\mbox{\small c}}}{\delta 
        j^a_{\bar{c}}} \Big) - \alpha f^{abc} j^b_b \,\frac{\delta 
        Z^{\mbox{\small c}}}{\delta j^c_{\bar{c}}} =  j^a_c 
\end{equation}
which, up to the undetermined coefficient $\alpha$, corresponds to the 
antighost equation (\ref{eomc}). In order to complete this identification, 
let us use the fact that, in any gauge theory with a linear gauge fixing 
condition, there exist a ghost equation which in our case follows from the 
Slavnov--Taylor identity differentiated with respect to the source for 
$n^\mu A_\mu$. Indeed this gives
\begin{equation} \label{ghost_z}
  n^\mu \partial_\mu \Big( \frac{\delta Z^{\mbox{\small c}}}{\delta j^a_c} 
           \Big) -
  f^{abc} j^b_b \,\frac{\delta Z^{\mbox{\small c}}}{\delta j^c_c} = 
         j^a_{\bar{c}} \quad ,
\end{equation}
which corresponds to (\ref{eomcb}).  
At this level, it is clear that consistency between (\ref{antighost_z}) 
and (\ref{ghost_z}) fixes the value $\alpha=1$.
Therefore, the equations of motion for the ghost sector 
(\ref{antighost_z}), 
(\ref{ghost_z}) are direct consequences of the symmetries. Furthermore, 
these equations show that the ghosts only couple to the source $J_b$ and 
therefore, it is possible to factorize out the effect of the ghost fields 
from the Slavnov--Taylor identity  (\ref{stid}). The latter is thus 
replaced by a local gauge Ward identity
\begin{eqnarray} \label{local_gauge}
\lefteqn{ -\partial_\mu j^{\m \, a} - n^\mu \partial_\mu 
        \Big( \frac{\delta Z^{\mbox{\small c}}}{\delta j^a_b} \Big) 
 +f^{abc} j^{\m \, b} \,\frac{\delta Z^{\mbox{\small c}}}{\delta j^{\m \, c}} 
 +f^{abc} j^b_\phi \,\frac{\delta Z^{\mbox{\small c}}}{\delta j^c_\phi} + 
     }\nonumber \\
&&     +\, f^{abc} j^b_c \,\frac{\delta Z^{\mbox{\small c}}}{\delta j^c_c} 
        + f^{abc} j^b_{\bar{c}} \,
        \frac{\delta Z^{\mbox{\small c}}}{\delta j^c_{\bar{c}}} 
        + f^{abc} j^b_b \,\frac{\delta 
        Z^{\mbox{\small c}}}{\delta j^c_b} = 0 \quad .
\end{eqnarray}

The Ward identity expressing the invariance of the theory under the 
residual gauge symmetry discussed in  sect. \ref{model} 
corresponds to the integration of (\ref{local_gauge}) with respect to
$x^1$. The residual Ward identity reads 
\begin{eqnarray} \label{gaugeresidu}
\lefteqn{\int_{-\infty}^{\infty} dx^1 \lac 
f^{abc} j^{\m \,  b} \,\frac{\delta Z^{\mbox{\small c}}}{\delta j^{\m\, c}} 
+f^{abc} j^b_\phi \,\frac{\delta Z^{\mbox{\small c}}}{\delta j^c_\phi} 
+\, f^{abc} j^b_c \,\frac{\delta Z^{\mbox{\small c}}}{\delta j^c_c}+
     \right. }\nonumber \\
&&      \left.   
     + f^{abc} j^b_{\bar{c}} \,
        \frac{\delta Z^{\mbox{\small c}}}{\delta j^c_{\bar{c}}} 
    + f^{abc} j^b_b \,\frac{\delta 
  Z^{\mbox{\small c}}}{\delta j^c_b}-\partial_2 j^{2\, a}\rac 
  =0  \quad .
\end{eqnarray}
It is important to notice that this step is plagued by the bad 
long distance behaviour of the field $b$. Indeed, passing from 
\equ{local_gauge} to \equ{gaugeresidu} would imply
\eq
\int_{-\infty}^{\infty} dx^1 \partial_1 
        \Big( \frac{\delta Z^{\mbox{\small c}}}{\delta j^a_b} \Big)=0
\eqn{irpro}
which is not the case as it can be shown using the
solutions given latter. This is the usual IR problem of the axial
gauge. In order to enforces \equ{irpro}, one substitutes  
\eq
\fud{}{J_b}\leftrightarrow e^{-\e(x^1)^2}\fud{}{J_b} \qquad (\e >0)
\eqn{bdamped}
which corresponds to a damping factor for the $b$--field along the $x^1$ 
direction, and takes the 
limit $\e\rightarrow 0$ at the end. It turns out that this simple substitution 
is sufficient in order to get \equ{irpro} and that the limit can be done 
trivially. The check is straightforward for all the Green's functions
and therefore is left to the reader. 

For the gauge sector, let us begin with the transversal
component  of the supersymmetry Ward identity and the ghost equation 
(\ref{ghost_z}) written as functional operator acting on
$Z^{\mbox{\small c}}$
\begin{eqnarray} 
\label{transversal_01}
        {\cal W}^{{\rm tr}} Z^{\mbox{\small c}}\equiv    
        {\cal W}_2 Z^{\mbox{\small c}} = 
        \int d^2x \,\left[ -\, j^a_\phi \,\frac{\delta 
        }{\delta j^a_{\bar{c}}} 
        - j^a_c \,\frac{\delta }{\delta j^{2\, a}} 
        - j^a_b \,\partial_2 
        \frac{\delta }{\delta j^a_{\bar{c}}} \right] Z^{\mbox{\small c}}
        & = & 0, \\
\label{gh_01} 
        {\cal G}^a Z^{\mbox{\small c}} = 
        \left( \partial_1 \frac{\delta }{\delta j^a_c}  -
        f^{abc} j^b_b \,\frac{\delta }{\delta j^c_c}  \right) 
           Z^{\mbox{\small c}} &=& j^a_{\bar{c}}.     
\end{eqnarray}       
Then, a direct calculation shows that the  consistency condition
\begin{equation} \label{wzcons}
\left\{ {\cal W}_2 , {\cal G}^a  \right\} Z^{\mbox{\small c}} 
                   = {\cal W}_2 j^a_{\bar{c}}  
\end{equation}
is in fact equivalent to the equation of motion for $A_2$ 
(\ref{eomf})
\begin{equation} \label{aeomsym}
\left( \pa_1  \frac{\delta }{\delta j^{2\, a}}  -
        f^{abc} j^b_b \,\frac{\delta }{\delta j^{2\, c}} \right)  
            Z^{\mbox{\small c}} =  j^a_\phi -\pa_2 j_b^a \quad .
\end{equation}

Up to now, we recover the equations of motion for $c$, $\bar{c}$ and 
$A_2$ as consistency conditions between the various symmetries but, 
contrary to the higher dimensional case, 
we did not get any information about the dynamic of $\phi$.

To clarify this point, let us consider the three--dimensional BF model 
\cite{bf3}. In this case, it is well known that the field $B$ is a 
one--form and therefore is invariant under the so--called reducible 
symmetry
\begin{equation}\label{redsym}
    s B^a_\mu  =  -\,(D_\mu \psi )^a
\end{equation}
where $\psi$ is a zero--form. To fix this extra symmetry one needs, 
besides the usual Yang-Mills ghost, 
antighost and Lagrange multiplier fields $(c, \bar{c}, b)$, a second set 
of such fields $(\psi, \bar{\psi}, d)$ related to (\ref{redsym}). The 
consequence  is 
that we exactly double the number of terms because of the exact similarity 
between the Yang-Mills part $(A_\mu , c , \bar{c} , b)$ and the reducible 
part $(B_\mu , \psi , \bar{\psi} , d)$. 
Therefore, it is easy to convince oneself that we recover the two 
antighost equations and, by considering the consistency 
conditions of the same type as (\ref{wzcons}), that we are in a position to 
construct the two 
equations of motion for $A_\mu$ and $B_\mu$\footnote{For a detailed 
discussion, see eqs. (4.3)--(4.5) in \cite{bf3}}.  For the 
two--dimensional case investigated here the field $\phi$ is a 0--form and 
does not exhibit any reducible symmetry of the type (\ref{redsym}). 
The invariance rather than covariance of $\phi$ is the basic difference 
between the two--dimensional BF model and  
the higher dimensional case. 

Finally, in order to clarify the present two--dimensional situation, let 
us collect all the functional identities we have got from symmetry 
requirements alone. These are the gauge condition (\ref{gauge_cond}), the 
antighost equation (\ref{antighost_z}) with $\alpha=1$, the ghost 
equation (\ref{ghost_z}), the transversal component of the 
supersymmetry Ward identity (\ref{transversal_01}), the field equation 
for $A_2$ (\ref{aeomsym}), the local Ward 
identity (\ref{local_gauge}) and the residual gauge Ward identity
(\ref{gaugeresidu}) which are respectively given  by\footnote{We will now 
systematically substitute \equ{axdef}.}
\begin{eqnarray}
\label{fgau}  
\frac{\delta Z^{\mbox{\small c}}}{\delta j^{1\, a}} 
                 & =&-\, j^a_b   \\
\label{fant} 
      \partial_1 \Big( \frac{\delta Z^{\mbox{\small c}}}{\delta 
      j^a_{\bar{c}}} \Big) -  f^{abc} j^b_b \,\frac{\delta 
      Z^{\mbox{\small c}}}{\delta j^c_{\bar{c}}}& =&  j^a_c   \\
\label{fgho} 
     \pa_1\Big( \frac{\delta Z^{\mbox{\small c}}}{\delta j^a_c} 
         \Big) -
  f^{abc} j^b_b \,\frac{\delta Z^{\mbox{\small c}}}{\delta j^c_c} 
                     &=& j^a_{\bar{c}}  \\
\label{fsutr}
\int d^2x \,\lac -\, j^a_\phi \,\frac{\delta 
        }{\delta j^a_{\bar{c}}} 
        - j^a_c \,\frac{\delta }{\delta j^{2\, a}} 
        - j^a_b \,\partial_2 
        \frac{\delta }{\delta j^a_{\bar{c}}} \rac Z^{\mbox{\small c}}
        & = & 0  \\
\label{fa}  
\left( \pa_1 \frac{\delta }{\delta j^{2\, a}}  -
        f^{abc} j^b_b \,\frac{\delta }{\delta j^{2\, c}} \right)  
            Z^{\mbox{\small c}} &= & j^a_\phi 
                -\pa_2  j^a_b,  \\
\label{floc}  
 \pa_1
\Big( \frac{\delta Z^{\mbox{\small c}}}{\delta j^a_b} \Big) 
+f^{abc}j^{\mu\, b}\,\frac{\delta Z^{\mbox{\small c}}}{\delta j^{\mu\, c}} 
 +f^{abc} j^b_\phi \,\frac{\delta Z^{\mbox{\small c}}}{\delta j^c_\phi} + 
         &&\nonumber \\
  +\, f^{abc} j^b_c \,\frac{\delta Z^{\mbox{\small c}}}{\delta j^c_c} 
   + f^{abc} j^b_{\bar{c}} \,
   \frac{\delta Z^{\mbox{\small c}}}{\delta j^c_{\bar{c}}} 
   + f^{abc} j^b_b \,\frac{\delta 
   Z^{\mbox{\small c}}}{\delta j^c_b} & =&  \partial_\mu j^{\m \, a} \\ 
\label{fres}
\int_{-\infty}^{\infty} dx^1 \lac 
f^{abc} j^{\m\, b} \,\frac{\delta Z^{\mbox{\small c}}}{\delta j^{\m\, c}} 
+f^{abc} j^b_\phi \,\frac{\delta Z^{\mbox{\small c}}}{\delta j^c_\phi} 
+\, f^{abc} j^b_c \,\frac{\delta Z^{\mbox{\small c}}}{\delta j^c_c} +
    \right.&&\nonumber \\
     \left.  
     + f^{abc} j^b_{\bar{c}} \,
        \frac{\delta Z^{\mbox{\small c}}}{\delta j^c_{\bar{c}}} 
    + f^{abc} j^b_b \,\frac{\delta 
  Z^{\mbox{\small c}}}{\delta j^c_b}-\pa_\m j^{\m \, a}\rac 
  & =&0  \quad .
\end{eqnarray}
As mentioned before, the procedure described above fails to produce an
 important relation. Indeed, the  equation of motion  
(\ref{eoma}) 
\begin{equation}\label{fphi}
\left( \varepsilon^{\mu\nu}\partial_\nu  \frac{\delta }{\delta j^a_\phi}  
    + n^\mu \frac{\delta }{\delta j^a_b}\right) Z^{\mbox{\small c}}
         + \varepsilon^{\mu\nu} f^{abc} 
         \frac{\delta Z^{\mbox{\small c}}}{\delta j^{\nu\, b}}
                 \frac{\delta Z^{\mbox{\small c}}}{\delta j^c_\phi}  
    -n^\mu f^{abc} \frac{\delta Z^{\mbox{\small c}}}{\delta j^b_c}
                 \frac{\delta Z^{\mbox{\small c}}}{\delta j^c_{\bar{c}}}
             = - j^{\mu\, a}
\end{equation}
is not a consequence of the symmetries and therefore, has to be derived 
from the action (\ref{total_s}).  More precisely, the $\m=1$ component 
of \equ{fphi} is the field equation for $\phi$ 
\eq
\left( \partial_1  \frac{\delta }{\delta j^a_\phi}  
         -f^{abc} j_b^b\frac{\delta }{\delta j^c_\phi}\rp 
         Z^{\mbox{\small c}}= - j^{2\, a}
\eqn{dynfi}
and the $\m=2$ component of \equ{fphi} is the field equation for $b$ 
\eq 
\lp \frac{\delta }{\delta j^a_b}- 
\partial_2 \frac{\delta }{\delta j^a_\phi}\rp Z^{\mbox{\small c}}            
   -f^{abc}\frac{\delta Z^{\mbox{\small c}}}{\delta j^{2\, b}}
                 \frac{\delta Z^{\mbox{\small c}}}{\delta j^c_\phi}  
    -f^{abc} \frac{\delta Z^{\mbox{\small c}}}{\delta j^b_c}
                 \frac{\delta Z^{\mbox{\small c}}}{\delta j^c_{\bar{c}}}
             = - j^{1\, a}
\eqn{intcond}
which is nothing else than an integrability condition. 

{\em Remarks:} 

All the functional identities obtained by symmetry 
considerations are linear in the quantum fields and, hence, they are not affected 
by radiative corrections. This linearization 
originates from the topological supersymmetry. 
 On the other hand, the equation for $b$ is quadratic and may causes problems.
This point will be treated later.

\section{Calculation of the Green's functions, perturbative finiteness} 
\label{prop_finit}
 
We will now derive the  solution to the set of equations
(\ref{fgau}) -- (\ref{fphi}). In turn, this will prove that the tree 
approximation (\ref{legendre}) corresponds to the exact solution.

\subsection{Solution of the Gauge Condition}
We already emphasized that the axial gauge allows for the factorization
of the ghost sector. This is illustrated by the solution for the gauge 
condition (\ref{fgau})  
\begin{equation}
\langle A^a_1(x) \, b^b(y) \rangle 
= -\,\delta^{ab} \delta^{(2)}(x-y) \quad .
\end{equation}        
which is the only non--vanishing Green's function containing  $A^a_1$.

\subsection{Solution of the ghost sector}
         
Let us first differentiate the antighost equation (\ref{fant})
with respect to the source $j^a_c$. This leads to\footnote{Our 
conventions for functional derivatives of even and/or odd objects are
$$
\fud{}{C}\int AB=\int \lp\fud{A}{C}B+(-1)^{{\rm deg}(A){\rm deg}(B)}
           \fud{B}{C}A\rp
$$} 
\begin{equation}
\pa_1 \frac{\delta^2 Z^{\mbox{\small c}}}{\delta j^a_{\bar{c}}(x) 
\,\delta j^b_c(y)} \Bigg|_{j = 0} = \,-\,\delta^{ab} \delta^{(2)}(x - y)
\end{equation}
A subsequent integration yields the propagator 
\eq
\langle \bar{c}^a(x) \, c^b(y) \rangle \, = 
\delta^{ab}\lc - \theta(x^1-y^1) \delta(x^2-y^2) + F(x^2-y^2) \rc
\eqn{cbcgen}
where $\theta$ is the step function,
\begin{equation}
        \theta(x-y) = \left\{ \begin{array}{ccl} 
                  1 & , & (x-y) > 0 \\
                  0 & , & (x-y) < 0 \quad . \end{array} \right.
\end{equation}
and  $F(x^2-y^2)$ is a function of  $x^2-y^2$ due to translational 
invariance, and with canonical dimension $1$. As the introduction of 
any dimensionfull parameters in our theory will spoil its topological 
character and that we work in the space of
tempered distributions\footnote{Any expression of the type 
 $1/(x^2-y^2)$ exhibits short distance singularities and 
therefore needs the introduction of a dimensionfull UV substraction point in 
order to give it a meaning.}\label{foutbol}, we are  left with   
\begin{equation} \label{cbarc}
\langle \bar{c}^a(x) \, c^b(y) \rangle \, =
-\,\delta^{ab} \lc \theta(x^1-y^1)+\a\rc \delta(x^2-y^2) 
\end{equation}
The analogous calculation starting form the ghost equation (\ref{fgho})
leads to the value 
\begin{equation} \label{einhalb}
        \alpha \; = \; -\,\frac{1}{2} \quad .
\end{equation}
for the integration constant as a consequence of the Fermi statistics
for the ghost fields  and the $c\leftrightarrow\bar{c}$ invariance 
of the theory. This implies the principal
value prescription for the unphysical pole $(n^\m k_\m)^{-1}$ in the 
Fourier--transform of the ghost--antighost propagator. 

It is important to note that contrarily to the Landau gauge 
\cite{blasi2}, the ghost--antighost propagator is infrared regular 
in the axial gauge. 

For the higher order Green's functions, the basic recurence relation is
obtained by differentiating (\ref{fant}) with respect to the most general
combination of the sources  
$(\delta^{(n+m+1)}/(\delta j_c)(\delta j_b)^n(\delta j_\varphi)^m)$ 
for $n+m\geq 1$ and  $\varphi\in \{A_2 , \phi , c, \bar{c}\}$. This
gives the following recursion relation over the number of $b$--fields 
\eqa \label{gen_rec}
\lefteqn{\pa_1\vev{\bar{c}^a(x) c^b(y)  b^{c_1}(z_1) 
\ldots b^{c_n}(z_n) 
\varphi^{d_1}(v_1) \ldots\varphi^{d_m}(v_m)} =}  \es
&&=\sum_{k = 1}^n f^{a c_k c}\,\d^{(2)}(x-z_k)\times \nes
&&\quad\times\vev{\bar{c}^c(z_k)  c^b(y) 
          b^{c_1}(z_1) \ldots \widehat{b^{c_k}(z_k)}
                \ldots b^{c_n}(z_n)  \varphi^{d_1}(v_1) \ldots 
                \varphi^{d_m}(v_m)} \nonumber
\eea
where $\widehat{\Phi}$ denotes the omission of the field $\Phi$ in 
the Green's functions. 

Let us first look at the case $n=0$ where (\ref{gen_rec}) reduces to
\eq 
\pa_1\vev{\bar{c}^a(x) \, c^b(y) \, 
\varphi^{d_1}(v_1) \ldots\varphi^{d_m}(v_m)}\; = 0
\eqn{gen_reczero}
 The solution is
\eq
\vev{\bar{c}^a(x) \, c^b(y) \, 
\varphi^{d_1}(v_1) \ldots\varphi^{d_m}(v_m)}\; = F(\x_k)
\qquad 1\leq k\leq M
\eqn{gensolzero}
where $\x_k$, stands for the $M=1+2m+\frac{m}{2}(m-1)$ differences
$\{x^2-y^2,\,x^2-v_i^2,\,v_i^2-v_j^2,\,y^2-v_i^2\}$, $1\leq i,j\leq m$
due to translational invariance. The absence of any 
dependance on the  coordinate 1 comes from the fact that any Green's 
function which does not involve $b$--fields obeys an homogeneous equation 
similar to (\ref{gen_reczero}) for all its arguments. 

Under the same assumptions as for (\ref{cbarc}),  $F(\x_k)$ has the general form
\eq
F(\x_k)\sim \d(\x_k)
\eqn{gensolfxi}
where the coefficients are either constants or proportional to 
$\ln(\frac{\x_k}{\x_{k'}})$ since this is the only combination which do not break 
scale invariance. Using now 
canonical dimension arguments (c.f. Tab. \ref{table1}),   conservation of 
the ghost charge
and residual gauge invariance \equ{fres}, one gets
\eq
\vev{(\bar{c})^{m_1} \, (c)^{m_2}\, (A_\mu)^{m_3}\, (\phi)^{m_4}}=0
\qquad \forall \ \{ m_1 ,\, m_2 ,\, m_3 ,\, m_4 \} \neq \{ 1,1,0,0 \}
\eqn{genzero} 

The next step  concerns the Green's functions which involves
$b$--fields. As a consequence of \equ{genzero}, the unique starting point 
for the recurence \equ{gen_rec} is  the two point function 
$\vev{c\, \bar{c}}$ \equ{cbarc}. Thus \equ{gen_rec} solves to the recurence 
relation 
\eqa \label{propghost}
\lefteqn{\vev{\bar{c}^a(x)  c^b(y)  b^{c_1}(z_1) \ldots 
                       b^{c_n}(z_n)} = } \es
     & & = \sum_{k = 1}^n f^{a c_k c} 
             [ \theta(x^1-z_k^1) + \alpha^{(n)} ] 
        \delta(x^2-z_k^2) \vev{\bar{c}^c(z_k)  
     c^b(y) b^{c_1}(z_1) 
      \ldots \widehat{b^{c_k}(z_k)} \ldots b^{c_n}(z_n)} \nonumber
\eea 
for $n\geq1$. 
The integration constants $\alpha^{(n)}$ are also fixed by the Fermi 
statistics of the ghost fields to be
\begin{equation}
        \alpha^{(n)} = -\,\frac{1}{2} \quad , \quad \forall \, n 
\end{equation}
and these solutions correspond to tree graphs 
\eqa\label{prop_barc_c_nb}
 \lefteqn{\langle \bar{c}^a(x) \, c^b(y) \, b^{c_1}(z_1) \,\ldots 
                      \, b^{c_n}(z_n) \rangle \; = } \es
     && = -\sum_{k = 1}^n f^{e c_k c} 
            \,\Big\langle\bar{c}^a(x)\, c^e(z_k)\Big\rangle
            \vev{\bar{c}^c(z_k) \, c^b(y) b^{c_1}(z_1) 
     \, \ldots \,\widehat{b^{c_k}(z_k)} \,\ldots b^{c_n}(z_n)} \nonumber
\eea
since (\ref{fant},\ref{fgho}) are linear in the quantum fields. Thus, this justify 
the tree approximation \equ{legendre} for this sector.

\subsection{Solution of the gauge sector}

Although we already know that the symmetries fail to produce
an important relation for this sector, let us look how 
far we can go in the determination of the Green's functions 
when taking into account only the symmetries for the model. 
The most fruitful approach is based on the transversal component of 
the supersymmetry Ward identity (\ref{fsutr}). 

The two points functions are found by differentiating (\ref{fsutr})
with respect to $\delta^{(2)}/\delta j^2 \delta j_c$, 
$\delta^{(2)}/\delta j_\phi \delta j_c$ and 
$\delta^{(2)}/\delta j_b \delta j_c$. They are         
\begin{eqnarray}\label{startaa}
       \langle A^a_2(y) \, A^b_2(z) \rangle & = & 
                      0 \quad , \\
       \label{prop_ccaphi} \langle A^a_2(x) \, \phi^b(y) \rangle 
           & = & \langle \bar{c}^b(y) \, c^a(x) \rangle \nonumber \\
            & = & -\,\delta^{ab} \,\lc\theta(y^1-x^1) - \frac{1}{2}\rc 
           \,\delta(x^2-y^2) \quad , \\
\label{startab}\langle b^a(x)\, A^b_2(y)  \rangle & = & \partial_2 
             \langle \bar{c}^a(x)\, c^b(y)  \rangle \nonumber \\
            & = & -\,\delta^{ab} \, \lc\theta(x^1-y^1) - \frac{1}{2}\rc \,
                \partial_2 \delta(x^2-y^2) \quad ,
\end{eqnarray}
where (\ref{cbarc}) and (\ref{einhalb}) have been used. 

 For the higher orders, 
(\ref{startaa},\, \ref{prop_ccaphi},\, \ref{startab}) generalize to 
\eqa
\vev{A^a_2(y) \, A^b_2(z)\, (\vf)^n}&=&0\nes
\vev{A^a_2(x) \, \phi^b(y)\, (\vf)^n}&=&
   \vev{\bar{c}^b(y) \, c^a(x)\, (\vf)^n}\nes
\vev{b^a(x)\, A^b_2(y)\, (\vf)^n}&=&
   \partial_2\vev{ \bar{c}^a(x)\, c^b(y)\, (\vf)^n}\nonumber
\eea
with  $\vf\in\{A_2,\phi,c,\bar{c},b\}$.
Since we have already the complete solution for the ghost sector 
(\ref{cbarc}\, ,\ref{genzero}\, \ref{prop_barc_c_nb}), this proves that in the 
axial gauge the supersymmetry completely fixes all the Green's functions 
which contain at least one field  $A_2$. 

\subsubsection{Solution of the Local Gauge Ward--identity}

The solution for $\langle (b)^n \rangle$, $\forall n$ can be derived 
from the local gauge Ward--identity (\ref{floc}). Indeed, by 
differentiation with respect to $\delta^{(n)}/(\delta j_b)^n$, $n\geq 1$,  
one gets directly
\begin{equation}
\langle (b)^{(n+1)} \rangle  =  0 \qquad \forall n\geq 1 \label{solbn}
\end{equation}
 Since these are the only loop diagrams of the theory, this shows that 
the non--linearity of \equ{intcond} have nn consequences and that the three 
approximation \equ{legendre} is exact.  

\subsubsection{Solution of the field equation for $\phi$}

 In the last two subsections, we showed that all the Green's function
of the form 
$$
\vev{A_2 \ldots}\quad , \quad \vev{b^n}
$$
where fixed by symmetry requirements. 
The remaining part formed by the Green's functions of the 
type $\langle (\phi)^m(b)^n \rangle$, $m\geq 1$ is solved only through the 
use of the equations of motion (\ref{fphi}). In the following, we 
will thus look for the general solution of (\ref{fphi}) for Green's functions 
with no  $A_2$ fields since the latter are already found in the previous 
subsection.

For the propagators, (\ref{fphi}) gives 
\begin{eqnarray}
\e^{\m\n}\pa_\n\vev{\phi^a(x) b^b(y) }&=&0\\
\partial_1 \langle \phi^a(x)\phi^b(y)\rangle&=&0 \\
\langle b^b(x) \phi^a(y) \rangle&=&
           \partial_2 \langle \phi^a(x)\phi^b(y)\rangle
\end{eqnarray}
which, together with (\ref{solbn}) and translational invariance solve into
\begin{eqnarray}
\langle \phi^a(x)\phi^b(y)\rangle&=& F(x^2-y^2) \label{fifi} \\
\langle b^b(x) \phi^a(y) \rangle&=&0  \label{bfi}
\end{eqnarray}
Here  $F(x^2-y^2)$ is an arbitrary function with canonical 
dimension $0$. Following the same reasoning as for (\ref{cbarc}), the 
latter is a constant
\eq
\vev{ \phi^a(x)\phi^b(y)}={\rm const}\;\d^{ab}
\eqn{vevphi}
It is important to notice that this constitute the first  
solution of the homogeneous equation which is not anihilated by the residual 
gauge invariance. This is caused by the bosonic character of the field
$\phi$ of canonical dimension $0$. 

The higher orders are generated by functional differentiating
 \equ{dynfi} with respect to  
$\delta^{(m+n)}/\delta (j_b)^m \delta (j_\phi)^n$
\begin{eqnarray}
\lefteqn{\partial_1\vev{\phi^a(x) 
       b(y_1)^{b_1}\ldots b(y_m)^{b_m} 
       \phi(z_1)^{c_1}\ldots\phi(z_n)^{c_n}}=}\label{genrecbf}\\[3mm]
&&=\sum_{i=1}^{m}f^{ab_ie}\delta^{(2)}(x-y_i) \vev{\phi^e(y_i) 
       b(y_1))^{b_1}\ldots \widehat{b(y_i))^{b_i}}\ldots b(y_m)^{b_m} 
       \phi(z_1)^{c_1}\ldots\phi(z_n)^{c_n} }\nonumber
\end{eqnarray}

For the $m=0$ case, the solution which generalizes \equ{fifi}
is 
\eq
F\lp\ln (\frac{x-z_i}{x-z_j}) \rp
\eqn{genso}
 but then \equ{intcond} imposes\footnote{See footnote ${}^6$ on 
p. \pageref{foutbol}.}
\eq
\vev{(\phi)^n}=\b_n \qquad \forall n
\eqn{vevgen}
where $\b_n$ is a constant which may depends on $n$.
Physically this correspond to the only invariants polynomials $\tr \phi^n$.

Furthermore, these solutions are the starting points for the reccurence 
\equ{genrecbf} for $m\neq 0$. Nevertheless, the Green's functions obtained
by this way do not satisfy the residual gauge invariance \equ{fres} and 
we must set $\b_n=0$.

\section{Conclusion}

 We already emphisize that the main difference of the two dimensional BF model
with respect to the higher dimensional cases is the absence of reducible 
symmetry caused by the 0--form nature of the field $\phi$.
The system is thus less constrained, \ie the symmetries do not fix all the 
Green's functions, the monomials $\tr(\phi)^n$ remains free.  

{\bf Acknowledgments:} One of the authors (S.E.) would like to thank the
``Fonds Turrettini'' and the ``Fonds F. Wurth'' for their financial support
during his stay at the Technische Universit\"at Wien where this work has 
been initiated. We are also indebted toward Olivier Piguet and Nicola Maggiore 
for helpful discussions.

\end{document}